\documentstyle[prd,preprint,aps,floats,epsfig]{revtex}
\draft
\newcommand{\beq}{\begin{equation}}
\newcommand{\eeq}{\end{equation}}
\newcommand{\bea}{\begin{eqnarray}}
\newcommand{\eea}{\end{eqnarray}}
\begin{document}

\title{On the extraction of skewed parton distributions from experiment} 

\author{Andreas Freund}

\address{I.N.F.N, Sezione Firenze, Largo Enrico Fermi 2, 50125 Firenze, Italy}

\maketitle
 
\begin{abstract}
In this paper we will discuss an algorithm for extracting skewed parton 
distributions from experiment as well as the relevant process and experimental
observable suitable for the extraction procedure.\newline
PACS: 12.38.Bx, 13.85.Fb, 13.85.Ni\newline
Keywords: Deeply Virtual Compton Scattering, Skewed Parton Distributions,\newline 
          Evolution, Factorization
\end{abstract}

\section{Introduction}
\label{intro}

Skewed parton distributions\footnote{This is the unified terminology since
the Regensburg conference of '98, finally eradicating the many terms like
non-diagonal, off-diagonal, non-forward and off-forward which have populated
the literature on this subject over the last few years. However recent 
publications have, alas, again fallen back upon the old terminology!} which 
appear in exclusive, hard diffractive processes like deeply virtual Compton 
scattering (DVCS)\footnote{First discussed in Ref.\ \cite{brod}.} 
or vector meson production with a rapidity gap, to name just
a few, have attracted a lot of theoretical and experimental interest over the
last few years as a hot bed for interesting new QCD physics \cite{1,2,3,4,5}. 
The list of references is
probably far from being complete and thus we apologize beforehand to everybody 
not mentioned.

\begin{figure}
\centering
\mbox{\epsfig{file=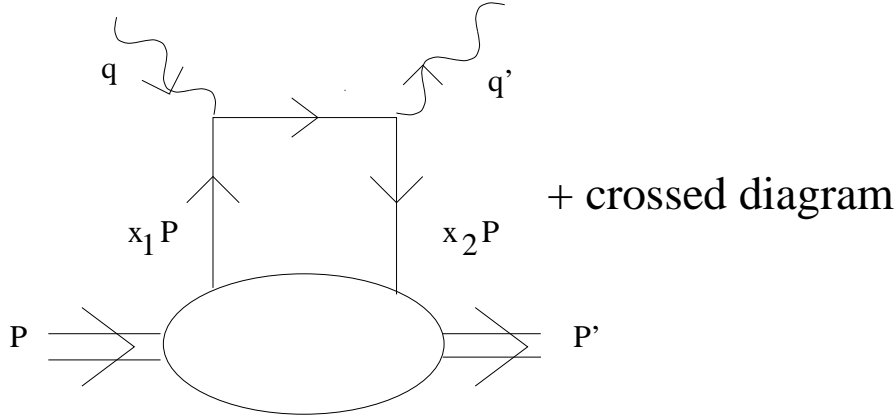,height=5.5cm}}
\caption{The lowest order handbag contribution to DVCS with $Q^2=-q^2$ and 
$q^{\prime 2}=0$.}
\label{hand}
\end{figure}

The basic concept of skewed parton distributions is illustrated in Fig.\ 
\ref{hand} with the lowest order graph of DVCS in which a quark of momentum 
fraction $x_1$ leaves the proton and is returned to the proton with momentum
fraction $x_2$. The two momentum fractions not being equal is due to the fact 
that an on-shell photon is produced which necessitates a change in the $+$ 
momentum in going from the virtual space-like photon with $+$ momentum usually 
taken to be $-xp_+$, where $p_+$ is the appropriate light cone momentum of the 
proton and $x$ is the usual Bjorken $x$, to basically zero $+$ momentum of the 
real photon. This sets $x_2=x_1-x$ and thus the skewedness parameter to $x$.
(see \cite{2} for more details on the kinematics.)

Thus one has a nonzero momentum transfer onto the proton and the parton 
distributions which enter the process are non longer the regular parton 
distributions of inclusive reactions since the matrix element of the appropriate
quark and gluon operators is now taken between states of unequal momentum rather
than equal momentum as in the inclusive case (see for example \cite{2}). These
parton distributions still obey DGLAP-type evolution equations but of a 
generalized form (see for example Radyushkin's references in \cite{1}).

The above mentioned kinematical situation is not the only one possible. One can also 
have the situation where $x_2$ becomes negative. In this case not a quark is 
returned to the proton but rather an anti-quark is emitted. In this situation one 
does not deal with parton distributions any more but rather distributional 
amplitudes obeying now Efremov-Radyushkin-Brodsky-LePage (ERBL) type evolution
equations (again see, for example, Radyushkin's references in \cite{1}). 
Furthermore, both momentum fractions could be negative in which case one is 
dealing with anti-quark distributions which again obey DGLAP-type evolution
equations.

After having answered the question how these skewed parton distributions arise,
the next question is which of the exclusive, hard diffractive processes is most suitable for 
extracting these skewed parton distributions and how can this be achieved. This
question will be answered in the following sections, where we discuss the most
promising process and the appropriate experimental observable in 
Sec.\ \ref{proc}, in Sec.\ \ref{extract} we will explain the algorithm and the 
problems associated with it and finally in Sec.\ \ref{concl} we will give an 
outlook on further research in this area.

\section{Appropriate Process and experimental observable}
\label{proc}

The most desirable process for extracting skewed parton distributions is the one
with the least theoretical uncertainty, the least singular $Q^2$ behavior so
as to be accessible over a broad range of $Q^2$ and 
with a proven factorization formula. The last requirement is actually the most 
important one since without a factorization theorem one has no reliable theoretical 
basis for extracting parton distributions.

The process which fulfills all the above criteria is DVCS since it is least
suppressed in $Q^2$ of all known exclusive, hard diffraction processes, in fact
it is only down by an additional factor of 
$Q^2$ in the differential cross section as compared to DIS\footnote{Compare this to 
the $1/Q^8$ 
behavior of vector meson production, di-muon production or di-jet production.},
the theoretical uncertainty is minimal since we are dealing with an elementary 
particle in the final state as compared to, for example, vector meson production 
where one also has to deal with the vector meson wavefunction in the factorization 
formula as an additional uncertainty and there exists a proven factorization 
formula \cite{2}.

Furthermore it was shown in Ref.\ \cite{3} that there will be sufficient DVCS
events at HERA as compared to DIS, albeit only at small $x$ between 
$10^{-4}-10^{-2}$, to allow an analysis with enough statistics.

The experimental observable which allows direct access to the skewed parton 
distributions is the azimuthal angle asymmetry $A$ of the combined DVCS and 
Bethe-Heitler(BH)\footnote{In the Bethe-Heitler process the incoming electron
exchanges a coulomb photon with the proton and radiates off a real photon, 
either before or after the interaction with the proton.} differential cross 
section, where the azimuthal angle is between the final state proton - $\gamma^*$
plane and the electron scattering plane. 
$A$ is defined as\cite{3}:
\beq
A = \frac{\int^{\pi/2}_{-\pi/2}d\phi~d\sigma_{DVCS+BH} - \int^{3\pi/2}_{\pi/2}
    d\phi~d\sigma_{DVCS+BH}}{\int^{2\pi}_{0}d\phi~d\sigma_{DVCS+BH}}.\\
\label{asym}
\eeq

In other words one counts the events where electron and photon are in the 
same hemisphere of the detector and 
subtracts the number of events where they are in opposite hemispheres
 and normalizes this expression with the total number of events.

The reason why this asymmetry is not $0$ is due to the interference term between
BH and DVCS which is proportional not only to $\cos (\phi)$, as compared to the 
pure DVCS and BH differential cross sections which are constant in $\phi$, but 
also to the real part of the DVCS amplitude. The factorized expression for the
real part of the amplitude takes the following form \cite{2}:
\beq
Re~T(x,Q^2) = \int^{1}_{-1+x}\frac{dy}{y} Re~C_{i}(x/y,Q^2)f_{i}(y,x,Q^2).\\
\label{factor}
\eeq
$Re~C_{i}$ is the real part of the hard scattering coefficient and $f_i$ are the
skewed parton distributions.  
The sum over the parton index $i$ is implied and $y$ is defined to be 
the parent momentum fraction in the parton distribution. 
As mentioned above, one is 
mainly restricted to the small-$x$ region where gluons dominate and thus $i$ in 
Eq.\ (\ref{factor}) will be only $g$ to a very good accuracy. Note that since the
parton distributions are purely real, the real part of the amplitude in its 
factorized form has to contain the real part of the hard scattering coefficient.

Thus Eq.\ (\ref{asym}) contains only measurable or directly computable quantities
except Eq.\ (\ref{factor}) in the interference part of the differential cross 
section for real photon production which is isolated in Eq.\ (\ref{asym}).
Therefore, one would now be
able to extract the skewed parton distributions from experimental information 
on $A$, the directly computable part of the interference term and the knowledge 
about the hard scattering coefficient if one could deconvolute 
Eq.\ (\ref{factor}). 

As we will see in the next section this direct deconvolution is not possible,
however there is a way around the deconvolution problem.

\section{Algorithms for extracting skewed parton distributions}
\label{extract}

\subsection{The Deconvolution Problem in DIS}

The deconvolution problem in inclusive DIS presents itself in a similar way as in
Eq.\ (\ref{factor}). For the structure function $F_2(x,Q^2)$, for example, one has 
the following factorization equation (in a general form):

\beq
F_2(x,Q^2) = \int^{1}_{x}\frac{dy}{y} C_{i}(x/y,Q^2)f_{i}(y,Q^2),\\
\label{factorf2}
\eeq
where one has the same situation as in Eq.\ (\ref{factor}) except, of course,
that the hard scattering coefficient $C_i$ and the parton distributions $f_i$ are
now different from the DVCS case. Also notice that the parton distributions depends
now only on $y$ rather than $y$ and $x$. In this case one can now easily
deconvolute Eq.\ (\ref{factorf2}) by taking moments in $x$ via 
$\int_{0}^{1}dx~x^N$. It is an easy exercise to show that the convolution integral
turns into a product in moment space:

\beq
\tilde F_2(N,Q^2) = \tilde C_{i}(N,Q^2) \tilde f_{i}(N,Q^2).\\
\label{factorf2N}
\eeq

Thus, after having calculated the hard scattering coefficient to the appropriate 
order and having measured $F_2$ such that the moment integral can be taken 
numerically, one can directly extract the parton distribution. What remains to be 
done is to perform the inverse Mellin transform to obtain the parton distribution
in terms of $x$ and $Q^2$. Of course, we have simplified the actual procedure and 
the inverse Mellin transform is also not easy to perform but this example serves
more as a pedagogical exercise to illustrate the basic concept of deconvolution
and extraction of parton distributions.

In the case of interest to us, however, life is not that "simple", since the
skewed parton distributions depend on two rather than one variable. Furthermore,
the hard scattering coefficient depends on the same variables as the parton
distribution. This makes the 
deconvolution of Eq.\ (\ref{factor}), at least to the best knowledge of the 
author, impossible because both the hard scattering coefficient and the parton 
distribution have two rather than one variable in common, one of which is even 
fixed, thus one does not have enough information to perform a deconvolution.

This seems like an intractable problem but there is a way out. For the purpose 
of as simple a presentation as possible the following discussion will only be
done in LO but the same principles also apply in NLO. However, the precision of 
the data in the foreseeable future, will be such that a leading order analysis 
will be sufficient. The following two discussions rest heavily on the methods in 
Ref.\ \cite{4,4a}. 

\subsection{The First Principle Extraction Algorithms}

The basic idea of this algorithm is to expand the parton distributions in terms
of orthogonal polynomials to reduce the unknown quantities in the factorization
formula for the real part of the DVCS amplitude to a number of unknown coefficients
which can be obtained through an inversion of a known matrix and DVCS data on the
real part of the amplitude.

As is well known, one can expand parton distributions or any smooth function 
for that matter, with respect to a complete set of orthogonal polynomials 
$P^{(\alpha_P)}_{j}(t)$, where $t$ is used here to shorten the notation. 
The orthogonality of the polynomials of our choice needs to be on the interval $-1 \leq t \leq 1$ with 
$t=\frac{2y-x}{2-x}$ which translates to an interval in $y$ of 
$-1+x\leq y \leq 1$ as found as the upper and lower bounds of the convolution 
integral in Eq.\ (\ref{factor}). One can then write the following expansion:
\beq
f^{q,g}(t,x,Q^2) = \frac{2}{2-x}\sum^{\infty}_{j=0}
\frac{w(t|\alpha_P)}{n_j(\alpha_P)}P_{j}^{q,g}(t)M^{q,g}_j(x,Q^2)
\label{expand}
\eeq
with $w(t|\alpha_P)$ and $n_j(\alpha_P)$ being weight and normalization 
factors determined by the choice of the orthogonal polynomial used. The labels $q,g$ 
for quarks and gluons are necessary since the $j$ label will be different 
for quarks and gluons. $\alpha_P$ is a label which depends on the orthogonal 
polynomials used\footnote{$\alpha_P =  \alpha, \beta$, in other words two 
labels, if Jacobi polynomials are used or $\alpha_P = \mu-1/2$ if Gegenbauer 
polynomials are used.}. $M^{q,g}_j(x,Q^2)$ is given by:
\beq
M^{q,g}_j(x,Q^2) = \sum^{\infty}_{k=0} E^{q,g}_{jk}(\nu ; \alpha_P |x)f^{q,g}_k(x,Q^2),
\label{coeff}
\eeq
where
\beq
f^{q,g}_k(x,Q^2) = \sum^{k}_{l=0} x^{l-k}B^{q,g}_{lk}\tilde f^{q,g}_l(x,Q^2).
\label{seceq}
\eeq
$B^{q,g}_{lk}$ is an operator transformation matrix which fixes the NLO 
corrections to the eigenfunctions of the kernels. The explicit form of the transformation matrix $B^{q,g}_{lk}$ 
can be found in Eq. $(35)$ of the second article of Ref.\ \cite{4} for example. The explicit 
form is not important, however, for neither this discussion nor the conclusion of this 
paper since we are dealing only with a LO analysis in which case the transformation matrix is just the 
identity matrix. The general from was just included for completeness sake. 

The moments $\tilde f^{q,g}_l(x,Q^2)$ of the parton distributions in Eq.\ (\ref{seceq}) generally evolve 
according to
\beq
\tilde f^{q,g}_l(x,Q^2) = \tilde K^{ik}_l (\alpha_s(Q^2),\alpha_s(Q^2_0))
\tilde f^{q,g}_l(x,Q^2_0)
\label{evol1}
\eeq
where the evolution operator is a matrix ($i$, $k$ equals either q or g) of functions in the 
singlet case 
(and just a function in the non-singlet case) taking account of 
quark and gluon mixing and depending on the order in the 
strong coupling constant.
 
Striving for simplicity, the above expansion is most simple in LO in the basis of Gegenbauer polynomials, since 
they are the eigenfunctions of the evolution kernels at LO. Thus we will use these polynomials from now on in our 
formulas. Thus, the Gegenbauer moments of the initial parton distributions at $Q_0^2$ in Eq.\ (\ref{evol1})
are defined the following way
\bea
\tilde f^{q}_{l}(x,Q_0^2) &=& \int^{1}_{-1}dt~\left ( \frac{x}{2-x} \right )^l C_l^{3/2}
\left ( \frac{tx}{2-x} \right ) f^{q}(t,x,Q^2_0)\nonumber\\
\tilde f^{g}_{l}(x,Q_0^2) &=& \int^{1}_{-1}dt~\left ( \frac{x}{2-x}\right )^{l-1}C_{l-1}^{5/2}
\left( \frac{tx}{2-x} \right ) f^{g}(t,x,Q^2_0).
\label{moments}
\eea 

Turning now to the expansion coefficients in Eq.\ (\ref{coeff}). 
The upper limit of the sum in Eq.\ (\ref{coeff}) is given by the constraint $\theta$-functions
\footnote{$\theta_{jk} = {1,~\mbox{if}~k\leq j;~0,~\mbox{if}~j<k}$} 
present in the expansion coefficients, which are defined, in terms of Gegenbauer polynomials, by
\bea
E^{q,g}_{jk}(\nu;\mu|x) &=& \frac{1}{2}\theta_{jk}\left [1+(-1)^{j-k}\right ]
\frac{\Gamma (\nu)}{\Gamma (\mu)}\frac{(-1)^{\frac{j-k}{2}}\Gamma (\mu + \frac{j+k}{2})}
{\Gamma (\nu + k) \Gamma (1 + \frac{j-k}{2})}\nonumber\\
& &(2-x)^{-k}~_2F_1 \left ( -\frac{j-k}{2}, \mu + \frac{j+k}{2}, \nu + k + 1|
\frac{x^2}{(2-x)^2}\right ),
\label{expcoef}
\eea
where $\nu = \mu = 3/2$ for quarks and $\nu = \mu = 5/2$ for gluons.

The general expansion in Eq.\ (\ref{expand}) then reduces to 
\bea
f^q(t,x,Q^2) &=& \frac{2}{2-x}\sum^{\infty}_{j=0}\sum^{\infty}_{k=0}\frac{w(t|3/2)}{N_j(3/2)}
E^{q}_{jk}(3/2|x) C_j^{3/2}(t)\tilde f^q_k(x,Q^2)\\
f^g(t,x,Q^2) &=& \frac{2}{2-x}\sum^{\infty}_{j=0}\sum^{\infty}_{k=1}\frac{w(t|5/2)}{N_j(5/2)}
E^{g}_{jk-1}(5/2|x)C_{j-1}^{5/2}(t)\tilde f^g_{k-1}(x,Q^2),
\label{expansion}
\eea
with $w(t|\nu) = (t(1-t))^{\nu -1/2}$, 
$N_j(\nu)=2^{-2\nu+1}\frac{\Gamma^2(1/2)\Gamma(2\nu+j)}{\Gamma^2(\nu)(\nu+j)j!}$ and 
the $C_j^{\nu}$ are Gegenbauer polynomials.

As we are in LO, multiplicatively renormalizable moments evolve with the following
explicit evolution operator:
\beq
\tilde K^{ik}_j(\alpha_s(Q^2),\alpha_s(Q^2_0)) = T exp\left( -\frac{1}{2}
\int^{Q^2}_{Q^2_0}\frac{d\tau}{\tau}\gamma^{ik}_j(\alpha_s(\tau)) \right )
\eeq
where $T$ orders the matrices of LO anomalous dimensions along the 
integration path. Note that there is a slight difference in the anomalous 
dimensions in the skewed case to the anomalous dimensions in the non-skewed, 
i.e. inclusive, case due to the particular definition of the conformal 
operators used in the definition of the parton distributions\footnote{
This is true in LO, in NLO, however, the anomalous dimensions obtain, besides
the NLO $\gamma_j$'s of the inclusive case, additional anomalous dimensions due
to non-diagonal elements in the renormalization matrix of the conformal operators
entering the skewed parton distributions (see Ref.\ \cite{5}).}  
$\gamma^{qg}_j = \frac{6}{j} \gamma^{qg,incl.}_j$ and $\gamma^{gq}_j = 
\frac{j}{6} \gamma^{gq,incl.}_j$.

Now we have all the ingredients to proceed. Inserting Eq.\ (\ref{expansion}) in
Eq.\ (\ref{factor}) one obtains for small $x$, where one is justified to neglect the quark contribution:
\bea
Re~T(x,Q^2) =& & 2\sum^{\infty}_{j=0}\sum^{\infty}_{k=1}
\tilde K^{gg}_{k-1} (\alpha_s(Q^2),\alpha_s(Q^2_0))\tilde f^{g}_{k-1}(x,Q^2_0)E^{g}_{jk-1}(5/2|x)
\nonumber\\
& &\int^{1}_{-1}\frac{dt}{2t+x}\frac{w(t|5/2)}{N_j(5/2)}Re~C_{g}\left ( 
\frac{1}{2}+\frac{t}{x},Q^2\right ) C_{j-1}^{5/2}(t),
\label{mastereq}
\eea
where we chose the factorization scale to be equal to the renormalization scale
, which we chose to be equal to $Q^2$.
As one can see the integral in the sum is now only over known functions and 
will yield, for fixed $x$, a function of $j$ as will also the expansion 
coefficients for fixed $x$. The evolution operator can also be evaluated and will yield 
for fixed $Q^2$ also just a function of $j$, which 
leaves the coefficients $\tilde f^{g}_{k-1}(x,Q^2_0)$ as the only unknowns, albeit
an infinite number of them. Since the lefthand side will be known from experiment
for fixed $x$ and $Q^2$, we are still in the unfortunate situation that a 
number is determined by the sum over an infinite number of coefficients labeled by $j$.
Thus, if one had measured the real part of the DVCS amplitude through the 
asymmetry $A$ at fixed $x$ \footnote{This fixes the undetermined coefficients up
to the $j$ index.} and at an infinite number of $Q^2$, one would have an infinite 
dimensional column vector on the lefthand side namely the real part of the 
amplitude at fixed $x$ but at an infinite number of $Q^2$ and on the right hand 
side one would have a square matrix\footnote{The number of $Q$ values determines
the column dimension and the number of the index $j$ determines the row 
dimension. The matrix is square since we can choose the number of $Q$ values to
be equal to the number of $j$ values!} times another column vector of 
coefficients of which the length is determined by the number of $j$. Since 
all the entries in the matrix are real and positive definite\footnote{The 
evolution operator will, of course, always yield a positive number, the 
integrals in the sum, are integrals over positive definite functions in the 
integration interval and the expansion coefficients are also positive definite
as can be seen from Eq.\ (\ref{expcoef}).}, it can be
inverted, using the well known linear algebra theorems on inversion of 
infinite dimensional square matrices, provided that there are no zero 
eigenvalues in other words no physical
zero modes in the problem which would imply that the real part of the DVCS 
amplitude would have to be zero which is, of course, never the case\footnote{
The DGLAP part of the amplitude will be zero at $x=1$ but the contribution 
from the ERBL region will not be!}. After having found the inverse, we can 
directly compute the moments of our initial parton distributions\footnote{Note
that the same moments of the initial parton distribution will appear for
different values of $j$, since the sum over $k$ runs up to $j$, for fixed $Q$,
such that each unknown moment is just multiplied by a number determined from 
known functions in Eq.\ (\ref{mastereq}).} which are needed to reconstruct the
skewed gluon distribution at small $x$ from Eq.\ (\ref{expand}).  

The drawback of the above procedure is that this process has to be repeated 
anew for each $x$. Nothing, however prevents us from doing so, in principle. 
Even for a finite number of $j$'s and $Q$'s, the task seems formidable, 
however, this is not as problematic as it seems, since experiment will only 
render information for small $x$, at least in the beginning, and not the whole 
range of $x$, thus one does not need an infinite number of coefficients and 
thus an infinite number of $Q$ for each $x$ to get a good approximation. 
Unfortunately a $j_{max}$ of $50-100$ will be necessary,
\footnote{The author's thanks go to Andrei Belitsky for pointing this out.}.
therefore, if the lefthand side is known for each $x$ at $50$ values of 
$Q^2$, Eq.\ (\ref{mastereq}) reduces to a system of $50$ 
equations with $50$ unknowns for $j_{max}=50$. This system can readily be 
solved as explained above. Experimentally speaking, of course, this procedure
is not feasible, though theoretically very attractive, since one will never be
able to measure the any experimental observable for fixed $x$ at $50$ different
values of $Q^2$. Nevertheless, there may be ways using constraints on SPD's to
reduce this number of $50-100$ polynomials as can be done in the forward case,
however this has to be further explored. 

Notwithstanding the above, let us give a toy example of the above extraction 
algorithm
\footnote{The author would like to thank John Collins 
for suggesting such an example to clarify the problem at hand.}. 
Take $x$ discrete and fix $Q^2$ then one can write
a factorized expression for a cross section:
\beq
\sigma_{a} = \sum_{j} H_{j;a} f_{j;a}.
\eeq
The index $j$ corresponds to the parton fractional momentum,
and $a$ to the $x$ variable.  Obviously, it is not possible to obtain $f_{j;a}$
from $\sigma_{a}$. If one now puts in an index for $Q$, the parton densities 
will now be $f(Q)_{j;a}$, and the solution of the evolution equation has the 
form 
\beq
f(Q)_{j;a} = \sum_k U(Q)_{j,k; a} f(Q_0)_{k; a}.
\eeq
Here $f(Q_0)_{k;a}$ is the initial parton density at the value $Q=Q_0$ which 
is left implicit. The cross section as a function of $Q$ takes now the form
\bea
\sigma_{Q;a} &=& \sum_{j,k} H_{j;a} U(Q)_{j,k; a} f(Q_0)_{k; a}\nonumber\\
              &=& \sum_k A(Q)_{j,k} f(Q_0)_{k;a},
\eea 
for a suitable matrix $A$. The $Q$ dependence of the hard scattering function 
$H$ can be ignored for our present purpose.

As a next step, take enough values of $Q$ such that the matrix is square. The 
most trivial example is to have two values of $Q$: the initial value
and one other:
\bea
           f_{1;j} &=& f(Q_0)_{j}\nonumber\\
           f_{2;j} &=& \sum_k U(Q)_{j,k} f(Q_0)_{k}.
\eea
One can take $U$ to be triangular, as is appropriate for DGLAP evolution
\beq
          U =  \left ( \matrix{1 & 1\cr 0 & 1} \right ),      
\eeq
the hard scattering cross section to be
\beq
          H = (1,1)
\eeq
and the parton distributions to be a two dimensional column vector:
\beq
         f_0 = \left ( \matrix{ f(Q_0)_{1} \cr f(Q_0)_{2}} \right ).
\eeq
This then yields
\beq
\left ( \matrix{\sigma(Q) \cr \sigma(Q_0)} \right) = \left ( \matrix{ 1 & 2 \cr
1 & 1} \right ) \left ( \matrix{f(Q_0)_{1} \cr f(Q_0)_{2}} \right ).\\
\label{final}
\eeq
Clearly one has an invertible matrix in Eq.\ (\ref{final}) and can thus compute
$f(Q_0)_{1}$ and $f(Q_0)_{2}$.

\subsection{The Practical Extraction Algorithm}

A practical way out of the polynomial predicament is by making a simple
minded ansatz for the skewed gluon distribution, since we are still at small
$x$, in the different ER-BL and DGLAP regions of the convolution integral Eq.\ 
(\ref{factor}). An example of such an Ansatz could be
$A_0z^{-A_1}(1-z)^A_3$ for the DGLAP region where $z$ is now just a dummy 
variable. If one inserts this Ansatz in Eq.\ (\ref{factor}) and can fit the 
coefficients to the data of the real part of the DVCS amplitude for fixed $x$ 
and $Q^2$. 
One can then repeat this procedure for different values of $Q^2$ and then interpolate
between the different coefficients to obtain a functional form of the 
coefficients in $Q^2$. Alternatively, after having extracted the values of
the coefficients for different values of $x$ at the same $Q^2$, use an evolution
programm with the ansatz and the fitted coefficients as input and check whether
one can reproduce the data for the real part at higher $Q^2$, thus checking
the viability of the model ansatz.

To obtain an ansatz fulfilling the various constraints for SPD's 
(see Ji's and Radyushkin's references in \cite{1,4a}), one should start from
the double distributions (DD) (see Redyushkin's 
references in \cite{1,4a}.) which yield the skewed gluon distribution
in the various regions of the convolution integral
\begin{eqnarray}
g(y,x) &=& \theta (y \geq x)\int^{\frac{1-y}{1-x}}_0 dz G(y-xz,z) +\nonumber\\
& & \theta (y \leq x) \int^{\frac{y}{x}}_0 dz G(y-xz,z).
\end{eqnarray}
Due to the fact that there are no anti-gluons, the above formula is enough to 
cover the whole region of interest $-1+x\leq y leq 1$. What remains is to choose
an appropriate model ansatz for G, for example,
\begin{equation}
G(z_1,z) = \frac{h(z_1,z)}{h(z_1)}f(z_1)
\end{equation}
with $f(z_1)$ being taken from a diagonal parametrization with its 
coefficients now being left as variants in the skewed case and
the normalization condition $h(z_1) = \int^{1-z1}_0 dz h(z_1,z)$ such that, in the
diagonal limit, the DD just gives the diagonal distribution. The choice for 
$h(z_1,z)$ is a matter of taste but should be kept as simple as possible. The
drawback of this algorithm as compared to the previous one is that it is model 
dependent and thus not a first principle methods, which theoretically speaking, is
not as satisfying but from the practical side this method is much simpler and 
thus experimentally much more feasible.

Thus, one has solved the problem of extracting the parton distributions from
the factorization equation, at least for small-$x$. The remaining problem is an
experimental one.

\section{Conclusions and outlook}
\label{concl}

After having showed, that the extraction of skewed parton distributions from
DVCS experiments is both principally and practically possible given the
high enough statistics data on the asymmetry, one should now get a more accurate
model description of the asymmetry. This will be done elsewhere.

\section*{Acknowledgments}

This work was supported by the E.\ U.\ contract $\#$FMRX-CT98-0194.

The author would like to thank John Collins, Andrei Belitsky and Mark Strikman
for helpful discussions and comments on the draft version of this paper.


\begin{references}

\bibitem{brod} S.J. Brodsky, F. Close and J.F. Gunion, Phys. Rev. {\bf D6}
               (1972) 177 and 2652.

\bibitem{1}    S.J. Brodsky, L.L. Frankfurt, J.F. Gunion, A.H. Mueller, 
               and M. Strikman, Phys. Rev. {\bf D50} (1994) 3134; 
               {\it ibid.} Erratum in Phys. Rev. D.\\
               A. Radyushkin Phys. Letters {\bf B385} (1996) 333,
                   Phys. Lett. {\bf B380} (1996) 417, Phys.\ Rev.\ {\bf D59}
               (1999) 014030 (hep-ph/9805342).\\
               J.C. Collins, L. Frankfurt, and M. Strikman, 
                    Phys. Rev. {\bf D56} (1997) 2982.\\
               X.-D. Ji, Phys. Rev. {\bf D55} (1997) 7114, Phys.\ Rev.\
                Lett.\ {\bf 78}, 610 (1997).\\
               L.L. Frankfurt, A. Freund, V. Guzey and M. Strikman,
               Phys.\ Lett.\ {\bf B 418}, 345 (1998), Erratum-ibid. {\bf B429} 
               (1998) 414\\
               A.Martin and M.Ryskin, Phys.\ Rev.\ {\bf D57}, 6692 (1998).\\
               A. Freund and V. Guzey, hep-ph/9801388 submitted to Comp.\
               Phys.\ Comm.\ and hep-ph/9806267 submitted to Phys.\ Let.\ B\\
               L. Mankiewicz, G. Piller and T. Weigel, Eur.\ Phys.\ J.\ {\bf C5} (1998) 119-128
               (hep-ph/9711227).\\
               X.-D. Ji and J. Osborne, Phys.\ Rev.\ {\bf D58} (1998) 094018 (hep-ph/9801260).\\
               D. M\"uller, Phys.\ Rev.\ {\bf D58} (1998) 054005 (hep-ph/9704406).\\
               X. Ji and J. Osborne, Phys.\ Rev.\ {\bf D57}, 1337  (1998).\\               
               A.V. Belitsky, B. Geyer, D. M\"uller, A. Sch\"afer
                Phys.\ Lett.\ {\bf B421}, 312 (1998).\\
               M. Diehl et al. , Phys.\ Lett.\ {\bf B411}, 193 (1997), 
               Phys.\ Lett.\ {\bf B428}, 359 (1998), Phys.\ Rev.\ {\bf D59} (1999) 034023
               (hep-ph/9808479) and Eur.\ Phys.\ J.\ {\bf C} (1999) DOI 
               http://dx.doi.org/10.1007/s100529901100 (hep-ph/9811253).\\
               Z. Chen, Nucl.\ Phys.\ {\bf B525} (1998) 369-383 (hep-ph/9705279.)\\       
               L. Mankiewicz, G. Piller, E. Stein, M. V\"attinen and T. Weigl,
               Phys.\ Lett.\ {\bf B425} (1998) 186-192 (hep-ph/9712251).\\
               J. Bl\"umlein, B. Geyer, and D. Robaschik, Phys.\ Lett.\ 
                {\bf B406}, 161 (1997), hep-ph/9711405.\\
               B. Pire, J. Soffer and O. V. Teryaev, hep-ph/9804284.\\               
               F.\ M.\ Dittes, J.\ Horejsi, B.\ Geyer, D.\ M\"uller and
               D.\ Robaschick, Phys.\ Lett.\ {\bf B209}, 325 (1988).\\
               A. Martin and K. Golec-Biernat, Phys.\ Rev.\ {\bf D59} (1999) 014029 
               (hep-ph/9807497).
               
\bibitem{2}    J.C. Collins and A. Freund, Phys.\ Rev.\ {\bf D59} (1999) 074009 (hep-ph/9801262)
               . A. Radyushkin, Phys.\ Rev.\ {\bf D56}, 5524 (1997).
               
\bibitem{3}    L. Frankfurt, A. Freund and M. Strikman, Phys.\ Rev.\ {\bf D58} (1998) 114001 
               (hep-ph/9710356) and hep-ph/9806535 submitted to Phys.\ Lett.\ B. 
               
\bibitem{4}    A.\ V.\ Belitsky, D.\ M\"uller, L.\ Niedermeier and A.\
               Sch\"afer, Phys.\ Lett.\ {\bf B437} (1998) 160-168 (hep-ph/9806232) and 
               hep-ph/9810275.
               
\bibitem{4a}   A. Radyushkin, Phys.\ Lett.\ {\bf B449} 81 (1999) (hep-ph/9810466).               

\bibitem{5}    A.V. Belitsky and D. M\"uller, Phys.\ Lett.\ {\bf B417}, 129
               (1998), Nucl.\ Phys.\ {\bf B527} (1998) 207-234 (hep-ph/9802411), 
               Nucl.\ Phys.\ {\bf B537} (1999) 397-442 (hep-ph/9804379).
                
\end{references}
\end{document}